\documentclass[12pt]{article}
\usepackage[english]{babel}
\usepackage{amsmath}
\usepackage{amsfonts}
\usepackage{amssymb}
\usepackage{hyperref}
\allowdisplaybreaks[1]

\begin{document}

\title{\textbf{On manifolds admitting the consistent Lagrangian formulation for
higher spin fields}}

\author{\sc I.L. Buchbinder${}^{a}$\thanks{joseph@tspu.edu.ru},
V.A. Krykhtin${}^{ab}$\thanks{krykhtin@tspu.edu.ru}, P.M.
Lavrov$^c$\thanks{lavrov@tspu.edu.ru}
\\[0.5cm]
\it ${}^a$Department of Theoretical Physics,\\
\it Tomsk State Pedagogical University,\\
\it Tomsk 634061, Russia\\[0.3cm]
\it ${}^b$Laboratory of Mathematical Physics,\\
\it Tomsk Polytechnic University,\\
\it Tomsk 634034, Russia\\[0.3cm]
\it ${}^c$Department of Mathematical Analysis,\\
\it Tomsk State Pedagogical University,\\
\it Tomsk 634061, Russia}
\date{}

\maketitle
\thispagestyle{empty}

\begin{abstract}
We study a possibility of Lagrangian formulation for free higher
spin bosonic totally symmetric tensor field on the background
manifold characterizing by the arbitrary metric, vector and third
rank tensor fields in framework of BRST approach. Assuming existence
of massless and flat limits in the Lagrangian and using the most
general form of the operators of constraints we show that the
algebra generated by these operators will be closed only for
constant curvature space with no nontrivial coupling to the third
rank tensor and the strength of the vector fields. This result
finally proves that the consistent Lagrangian formulation at the
conditions under consideration is possible only in constant
curvature Riemann space.
\end{abstract}


Lagrangian formulation of interacting higher spin field theory is a
fundamental unsolved problem of classical field theory during long
time (see e.g. the reviews \cite{reviews}). The essence of the
problem is that any naive (e.g. minimal) including the interaction
to free higher spin field Lagrangian violates consistency of the
equations of motion (see the various aspects of the inconsistency in
\cite{consist,consist1,consist2}). One of the partial aspects of the
generic problem is a Lagrangian formulation for higher spin fields
coupled to external background. At present, all known consistent
Lagrangian formulations are constructed only on space of constant
curvature without any other external fields. Then a natural question
arises if there exist the other background fields admitting the
consistent Lagrangian formulation for higher spin fields. For
example, if to accept that the massive higher spin fields have the
superstring origin, the evident background in bosonic sector is
formed by the fields from massless string spectrum what corresponds
in sigma-model approach \cite{FT} to manifold endowed with Riemann
metric and additional external scalar (dilaton), vector and totally
antisymmetric third rank tensor field which can be associated with
torsion. In principle one can hope that the consistent Lagrangian
formulation of higher spin fields coupled to background fields
actually exists under some equations linking all the background
fields.

In this note we consider the free massive higher spin bosonic field
theory coupled to external metric, vector field and arbitrary third
rank tensor field and assume that the Lagrangian contains no inverse
mass terms. At these conditions we prove that the only manifold
admitting the consistent Lagrangian formulation is constant
curvature space with vanishing scalar, vector and third rank tensor
external fields. The proof is based on generic BRST formulation of
higher spin field theory (see the various use of BRST formalism in
higher spin field theory in
\cite{BRSThsf,0611,bkl,brst1,BRSThsf1,mg}
which allows efficiently
to study the bosonic and fermionic, massless and massive higher spin
fields, to take into account a gauge structure of the theory and to
work with tensor fields of various symmetry of indices.

The BRST approach to Lagrangian construction for higher spin fields,
which is developed in our papers, is realized as follows. The
mass-shell equation and subsidiary conditions are treated as a part
of the first class constraints of some unknown yet gauge theory. The
new constraints are added to initial ones to form a complete set of
first class constraints. All the constraints are formulated as the
operators acting in auxiliary Fock space and it is assumed that the
algebra of the constraints in terms of commutators is closed. Taking
into account these constraints one can construct the Hermitian
nilpotent BRST-BFV operator $Q$ \cite{BF} and find higher spin field
Lagrangian in terms of the $Q$. Nilpotency of the $Q$ guarantees a
gauge invariance of the corresponding action. As a result we get the
higher spin field Lagrangian which automatically contains all the
auxiliary fields. Such an approach has been completely realized in
$d$-dimensional flat and AdS spaces and its different aspects have
been studied in \cite{BRSThsf,0611,bkl,BRSThsf1}. The essential
basic element of such approach was closure of the algebra of
operator constraints.

To realize this approach for the fields on some background
first of all we should find the corresponding mass-shell and subsidiary
conditions. For bosonic fields in flat space they are Klein-Gordon
equation and the conditions for the fields to be divergence free and
traceless (see e.g. \cite{bk})
\begin{eqnarray}
(\partial^2-m^2)\varphi_{\mu_1\ldots\mu_s}=0,
\qquad
\partial^{\mu_1}\varphi_{\mu_1\ldots\mu_s}=0,
\qquad
\eta^{\mu_1\mu_2}\varphi_{\mu_1\mu_2\ldots\mu_s}=0.
\label{EM-flat}
\end{eqnarray}
In case of fields on a background all the conditions (\ref{EM-flat})
should be deformed by the proper way. For the bosonic fields in AdS
space such deformations has been constructed in \cite{bkl}. It is
interesting to point out that in the case of fields in AdS space the
corresponding closed algebra of constraints belongs to class of
quadratic non-linear algebra and finding the nilpotent BRST operator
is a very nontrivial problem (see e.g. \cite{NvH}). In the case
under consideration we should construct a deformation of the
constraints (\ref{EM-flat}), realizing all the conditions as the
operator constraints and find the restrictions on the background
when their algebra will be closed.

We assume that a manifold under consideration is endowed with metric
$g_{\mu\nu}$, the background vector $A_{\mu}$ and third rank tensor
$K_{\mu\nu\alpha}$ field with no index symmetry, so that the torsion
tensor is a particular case of $K_{\mu\nu\alpha}$ tensor.
The covariant derivatives are constructed with the use of
Christoffel symbols $
\Gamma^\alpha_{\mu\nu}=\frac{1}{2}g^{\alpha\beta} (\partial_\mu
g_{\beta\nu}+\partial_\nu g_{\mu\beta}-\partial_\beta g_{\mu\nu}). $


Introduce the auxiliary Fock space generated by bosonic creation and
annihilation operators with tangent space indices
($a,b=0,1,\ldots,d-1$)
\begin{eqnarray}
[a_a,a^+_b]=\eta_{ab},
\qquad
\eta_{ab}=diag(-,+,\ldots,+).
\label{a-bos}
\end{eqnarray}
An arbitrary vector in this Fock space has the form
\begin{eqnarray}
\label{PhysState}
|\varphi\rangle
&=&
\sum_{s=0}^{\infty}\varphi_{a_1\ldots\,a_s}(x)\,
a^{+a_1}\ldots\,a^{+a_s}|0\rangle
=
\sum_{s=0}^{\infty}\varphi_{\mu_1\ldots\,\mu_s}(x)\,
a^{+\mu_1}\ldots\,a^{+\mu_s}|0\rangle
\equiv
\sum_{s=0}^{\infty}|\varphi_s\rangle
,
\end{eqnarray}
where
$a^{+\mu}(x)=e^\mu_a(x)a^{+a}$, $a^\mu(x)=e^\mu_a(x)a^a$,
with $e^\mu_a(x)$ being the vielbein.
It is evident that $[a_\mu, a_\nu^+]=g_{\mu\nu}$.
We also
suppose the standard relation
$\nabla_\mu e^a_\nu=\partial_\mu e^a_\nu-\Gamma^\alpha{}_{\nu\mu}e^a_\alpha
+\omega_\mu{}^a{}_be^b_\nu=0$, where $\omega_\mu{}^a{}_b$ is the spin connection.

Then one
introduces derivative operator
\begin{eqnarray}
D_\mu=\partial_\mu+\omega_\mu^{ab}a_a^+a_b, &\qquad& D_\mu|0\rangle=0
\label{D_mu}
\end{eqnarray}
which acts on states of the Fock space (\ref{PhysState}) as the covariant
derivative
\begin{eqnarray}
D_\mu|\varphi_s\rangle=(\nabla_\mu\varphi_{\mu_1\ldots\,\mu_s})\,
a^{+\mu_1}\ldots\,a^{+\mu_s}|0\rangle
\end{eqnarray}
and tries to realize the generalization of equations (\ref{EM-flat}) in the operator
form
\begin{eqnarray}
l_0|\varphi_s\rangle=
l_1|\varphi_s\rangle=
l_2|\varphi_s\rangle=
0
\label{EM-op}
\end{eqnarray}
where operators $l_0$, $l_1$, $l_2$ corresponding to Klein-Gordon,
divergence free and traceless equations respectively.

The procedure of Lagrangian construction based on the BRST method looks as follows.
For the Lagrangian be a real function the BRST operator used for its construction must
be a Hermitian operator. It assumes that the set of operators underlying the BRST
operator must be invariant under Hermitian conjugation.
To have such a set of operators we add to constraints $l_0$, $l_1$, $l_2$ their
Hermitian conjugated operators $l_1^+$, $l_2^+$
with $l_0$ being assumed to be self-conjugated.
Then for constructing the BRST operator the underlying set of
operators must form an algebra.
To get the algebra we must add to operators $l_0$, $l_1$, $l_2$, $l_1^+$, $l_2^+$ some more operators providing closing the algebra.
But if we want to construct with the help of the obtained algebra
Lagrangian for spin-s field, then this algebra must be a deformation
of the algebra in Minkowski or in AdS \cite{bkl}
space. Thus we can add only two operators which are generalization
of operators
\begin{eqnarray}
g_0=a_\mu^+a^\mu+\tfrac{d}{2},
\qquad
g_m=m^2+const
\label{g0}
\end{eqnarray}
to the case of curved space. Since operator $g_0$ is dimensionless
and we do not consider terms with inverse powers of the mass then it
is impossible to deform operator $g_0$ by terms with the curvature or with
the background fields.\footnote{We adopt that the backgound fields
have mass dimension one like the torsion.}
Therefore operator $g_0$ keeps the same form (\ref{g0}) as in the
flat case. As for possible generalization of operator $g_m$ we postpone this question.
Thus we came to the conclusion that
in
order to be possible to construct Lagrangian with the help of the BRST method
we must find explicit expressions for the operators
\begin{eqnarray}
l_0,\qquad l_1,\qquad l_1^+,\qquad l_2,\qquad l_2^+,\qquad g_0,\qquad g_m
\label{set-R}
\end{eqnarray}
so that they form an algebra.\footnote{\label{fnote}It should be note that in general case, operators $l_0$, $l_1$, $l_2$
may not coincide with that given in (\ref{EM-op}).
We demand here only that operators (\ref{set-R}) have the proper free limit.
The consistent conditions on the field (\ref{EM-op}) will be followed from the BRST construction.}
If we find algebra different from the AdS case,
then it means that there is a hope to construct Lagrangian in space
different from AdS.

We will deform the operators by introducing background fields.
We explore the case when the background fields are vector and third rank
tensor with dimension of mass.
Note that any third rank tensor can be decomposed into totally
symmetric $S_{(\mu\nu\sigma)}$, totally antisymmetric $A_{[\mu\nu\sigma]}$
tensors, and tensors with mixed symmetry of the indices (see e.g.
\cite{0611263}).
We consider the case of the decomposition when the mixed symmetry tensors
have the following symmetry of the indices
\begin{eqnarray}
\label{M-sym}
M_{\mu\nu\sigma}=-M_{[\nu\mu]\sigma}\,,
&&
M_{[\mu\nu\sigma]}=0\,.
\end{eqnarray}
In addition we adopt that all the background tensors are traceless,
absorbing their traces into vector field $V_\mu$.

Let us discuss possible form of the operators.
First, the dimensionless operators
\begin{eqnarray}
 \label{l2}
l_2={\textstyle\frac{1}{2}}\,a^\mu a_\mu
\,,
\qquad
l_2^+=\tfrac{1}{2}a^+_\mu a^{\mu+}
\,,
\qquad
g_0=a_\mu^+a^\mu+\tfrac{d}{2}
\end{eqnarray}
can't be modified by the background fields if we don't take into account the mass in the inverse powers.

Next let us consider operator $l_1$ responsible for the physical
field to be divergence free. Since this operator has mass dimension
one then the background fields are introduced linearly in $l_1$.
Also we note that in the terms with the background fields the
creation and annihilation operators, if they not contracted with the
background fields, are contracted with each other and these
contractions can be expressed through the operators $l_2$, $l_2^+$,
$g_0$ (\ref{l2}). Therefore the most general expression for the
operator $l_1$ is
\begin{eqnarray}
l_1&=&a^\mu D_\mu
+\sum_{k,m=0}^{\infty}\alpha_{km}\; V_\mu a^\mu\;(l_2^+)^m \; g_0^k  \; l_2^m
+\sum_{k,m=0}^{\infty}\omega_{km}\; V_\mu a^{+\mu}\;(l_2^+)^m \; g_0^k \; l_2^{m+1}
\nonumber
\\
&&{}+\sum_{k,m=0}^{\infty}\beta_{km}\; M_{[\mu\nu]\sigma} a^{+\mu}a^\nu a^\sigma\;
(l_2^+)^m \; g_0^k \; l_2^m
+\sum_{k,m=0}^{\infty}\sigma_{km}\; M_{[\mu\nu]\sigma}
a^{+\nu}a^{+\sigma}a^\mu\;(l_2^+)^m \; g_0^k  \; l_2^{m+1}
\nonumber
\\
&&{}+\sum_{k,m=0}^{\infty}\epsilon_{km}\;S_{(\mu\nu\sigma)}a^{+\mu} a^{+\nu} a^{+\sigma}
\;(l_2^+)^m \; g_0^k \; l_2^{m+2}
+\sum_{k,m=0}^{\infty}\zeta_{km}\;S_{(\mu\nu\sigma)}a^{+\mu} a^{+\nu}a^{\sigma}
\;(l_2^+)^m \; g_0^k  \; l_2^{m+1}
\nonumber
\\
&&{}+\sum_{k,m=0}^{\infty}\gamma_{km}\;S_{(\mu\nu\sigma)}a^{+\mu}a^{\nu}a^{\sigma}
\;(l_2^+)^m \; g_0^k \; l_2^m
+\sum_{k,m=0}^{\infty}\theta_{km}\;S_{(\mu\nu\sigma)}a^{\mu}a^{\nu}a^{\sigma}
\;(l_2^+)^{m+1} \; g_0^k \; l_2^m
\label{l1}
\end{eqnarray}
with arbitrary coefficients
$\alpha_{km}$, $\beta_{km}$, $\gamma_{km}$, $\omega_{km}$,
$\sigma_{km}$, $\epsilon_{km}$, $\zeta_{km}$, $\theta_{km}$.
The totally antisymmetric tensor $A_{[\mu\nu\sigma]}$ cannot be introduced
into $l_1$ since any its contraction with creation and annihilation operators
gives zero. But it should be noted that in case we considered the dynamics of a field
with mixed symmetry of the indices in the background fields then the totally
antisymmetric field $A_{[\mu\nu\sigma]}$ could be introduced into $l_1$.

Taking Hermitian conjugation (\ref{l1}) and moving operators
$l_2^+$, $g_0$, $l_2$
to the right we obtain expression for the operator $l_1^+$
\begin{eqnarray}
l_1^+
&=&
-a^{+\mu} D_\mu
+\sum_{k,m=0}^{\infty}\alpha'_{km}\;V_\mu a^{+\mu}\;(l_2^+)^m \; g_0^k\;l_2^m
+\sum_{k,m=0}^{\infty}\omega'_{km}\;V_\mu a^\mu \;(l_2^+)^{m+1}\;g_0^k\;l_2^m
\nonumber
\\
&&{}+\sum_{k,m=0}^{\infty}\beta'_{km}\;M_{[\mu\nu]\sigma}a^{+\nu}a^{+\sigma}a^\mu
\;(l_2^+)^m \; g_0^k \; l_2^m
+\sum_{k,m=0}^{\infty}\sigma'_{km}\;M_{[\mu\nu]\sigma}a^{+\mu}a^\nu a^\sigma
\;(l_2^+)^{m+1} \; g_0^k \; l_2^m
\nonumber
\\
&&{}
+\sum_{k,m=0}^{\infty}\epsilon'_{km}\;S_{(\mu\nu\sigma)}a^\mu a^\nu a^\sigma
\;(l_2^+)^{m+2} \; g_0^k \; l_2^m
+\sum_{k,m=0}^{\infty}\zeta'_{km}\;S_{(\mu\nu\sigma)}a^{+\mu}a^\nu a^\sigma
\;(l_2^+)^{m+1} \; g_0^k \; l_2^m
\nonumber
\\
&&{}
+\sum_{k,m=0}^{\infty}\gamma'_{km}\;S_{(\mu\nu\sigma)}a^{+\mu}a^{+\nu}a^{\sigma}
\;(l_2^+)^m \; g_0^k \; l_2^m
\nonumber
\\
&&{}
+\sum_{k,m=0}^{\infty}\theta'_{km}\;S_{(\mu\nu\sigma)}a^{+\mu}a^{+\nu}a^{+\sigma}
\;(l_2^+)^m \; g_0^k \; l_2^{m+1}
\label{l1+}
\end{eqnarray}
where the primed coefficients can be expressed through non-primed
ones and vice versa.
Moreover the dependent and independent coefficients can be chosen in a variety
of ways, choosing as independent coefficients partially both primed and non-primed coefficients.
Note that the terms containing at least one of
the operators $l_2^+$, $g_0$, $l_2$ don't influence on closing the
algebra (and as a consequence on the background geometry) and this
fact we denote as follows
\begin{eqnarray}
l_1&\approx&a^\mu D_\mu
+\alpha_{00}\; V_\mu a^\mu
+\beta_{00}\; M_{[\mu\nu]\sigma} a^{+\mu}a^\nu a^\sigma
+\gamma_{00}\;S_{(\mu\nu\sigma)}a^{+\mu}a^{\nu}a^{\sigma}
\label{l1a}
\\
l_1^+
&\approx&
-a^{+\mu} D_\mu
+\alpha'_{00}\;V_\mu a^{+\mu}
+\beta'_{00}\;M_{[\mu\nu]\sigma}a^{+\nu}a^{+\sigma}a^\mu
+\gamma'_{00}\;S_{(\mu\nu\sigma)}a^{+\mu}a^{+\nu}a^{\sigma}
\label{l1+a}
\end{eqnarray}
where $\approx$ means ``up to terms proportional to operators $l_2^+$, $g_0$, $l_2$''.
Also we note that $\alpha_{00}$, $\alpha_{00}'$, $\beta_{00}$,
$\beta_{00}'$, $\gamma_{00}$, $\gamma'_{00}$ can be considered as
independent of each other.

Let us consider commutators
\begin{eqnarray}
[\,l_1,l_2]\approx\gamma_{00}\; S_{(\mu\nu\sigma)}a^{\mu} a^{\nu} a^{\sigma}
\,,
&\qquad&
[\,l_1^+,l_2^+]\approx\gamma'_{00}\; S_{(\mu\nu\sigma)}a^{+\mu} a^{+\nu} a^{+\sigma}
\,.
\end{eqnarray}
We see that to close the algebra we must demand $\gamma_{00}=\gamma'_{00}=0$.
This means that the totally symmetric tensor $S_{(\mu\nu\sigma)}$ cannot
influence on the background geometry.

Next we consider commutators
\begin{eqnarray}
[\,l_1,l_2^+] &\approx& a^{+\mu}D_\mu+\alpha_{00}\; V_\mu a^{+\mu}
-\beta_{00} \; M_{[\mu\nu]\sigma} a^{+\nu}a^{+\sigma}a^\mu
\nonumber
\\
&\approx&-l_1^++(\alpha_{00}+\alpha'_{00})\; V_\mu a^{+\mu}
+(\beta'_{00}-\beta_{00}) \; M_{[\mu\nu]\sigma} a^{+\nu}a^{+\sigma}a^\mu
\,,
\\
{}
[l_1^+,l_2]&\approx&
l_1-(\alpha_{00}+\alpha'_{00})\; V_\mu a^{\mu}
+(\beta'_{00}-\beta_{00}) \; M_{[\mu\nu]\sigma}a^{+\mu}a^{\nu}a^{\sigma}
\,,
\end{eqnarray}
and for their closing it is necessary to put $\alpha'_{00}=-\alpha_{00}$
and $\beta'_{00}=\beta_{00}$.
Thus operators $l_1$ and $l_1^+$
will take the form
\begin{eqnarray}
l_1&\approx&a^\mu D_\mu+\alpha_{00}V_\mu a^\mu+\beta_{00}M_{[\mu\nu]\sigma} a^{+\mu}a^\nu a^\sigma
\label{l1b}
\\
l_1^+
&\approx&
-a^{+\mu} D_\mu-\alpha_{00}V_\mu a^{+\mu}+\beta_{00}M_{[\mu\nu]\sigma}a^{+\nu}a^{+\sigma}a^\mu
\label{l1+b}
\end{eqnarray}

Let us now consider commutator
\begin{eqnarray}
[l_1^+,l_1]&\sim&
D^2
+P^{\mu\alpha\sigma}a^+_\mu a_\alpha D_\sigma
+2\alpha_{00}V^\sigma D_\sigma
+W^{\mu\nu\alpha\beta}a^+_\mu a^+_\nu a_\alpha a_\beta
+K^{\mu\alpha}a^+_\mu a_\alpha
+Z
\,,
\label{l1+l1}
\end{eqnarray}
where
\begin{eqnarray}
&&
D^2=g^{\mu\nu}(D_\mu D_\nu-\Gamma^\sigma{}_{\mu\nu}D_\sigma)\,,
\\
&&
\label{P}
P^{\mu\alpha\sigma}=2\beta_{00}(M^{\mu(\alpha\sigma)}-M^{\alpha(\mu\sigma)})
=-P^{\alpha\mu\sigma}\,,
\\
&&
W_{\mu\nu\alpha\beta}=R_{\mu\alpha\beta\nu}
-\beta_{00}\bigl[\nabla_{(\beta} M_{\alpha)\mu\nu}+\nabla_{(\mu} M_{\nu)\alpha\beta}\bigr]
\nonumber
\\
&&
\mspace{82mu}{}
+\beta_{00}^2\bigl[M_{\tau(\mu\nu)}M^\tau{}_{(\alpha\beta)}
-4M_{\alpha(\tau\mu)}M_{\nu(\beta}{}^{\tau)}\bigr]
\,,
\\
&&
K_{\mu\alpha}=R_{\mu\alpha}
-2\beta_{00}\nabla^\sigma M_{\alpha(\mu\sigma)}
-2\beta_{00}^2M_{\mu(\rho\tau)}M_\alpha{}^{\rho\tau}
\nonumber
\\
&&
\mspace{82mu}{}
+\alpha_{00}(\nabla_\alpha V_\mu-\nabla_\mu V_\alpha)
+2\alpha_{00}\beta_{00}(M_{\mu(\alpha\sigma)}-M_{\alpha(\mu\sigma)})V^\sigma
\,,
\\
&&
Z=\alpha_{00}^2V_\mu V^\mu+\alpha_{00}\nabla_\mu V^\mu
\end{eqnarray}
and $\sim$ means ``up to terms proportional to operators
$l_1$, $l_1^+$, $l_2$, $l_2^+$, $g_0$''.
In order to have a closed algebra we have to suppose that the right hand side
of (\ref{l1+l1}) be proportional to operators of the algebra (\ref{set-R}).
For example we may define operators $l_0$ and $g_m$ as follows
\begin{eqnarray}
&&
l_0\sim
D^2-m^2
+P^{\mu\alpha\sigma}a^+_\mu a_\alpha D_\sigma
+2\alpha_{00}V^\sigma D_\sigma
+W^{\mu\nu\alpha\beta}a^+_\mu a^+_\nu a_\alpha a_\beta
+K^{\mu\alpha}a^+_\mu a_\alpha
+Z
\,,
\label{l0}
\\
&&
g_m=m^2
\,.
\end{eqnarray}

Let us turn to the commutator $[l_1^+,l_0]$ and consider terms with two
derivative operators. One has
\begin{eqnarray}
[l_1^+,l_0]&\sim&P^{\alpha(\mu\sigma)}a_\alpha^+ D_{(\mu}D_{\sigma)}+\ldots
\end{eqnarray}
Demanding that this commutator
be proportional to operators (\ref{set-R})
and since we are working with traceless fields, then  one has to suppose
\begin{eqnarray}
P^{\alpha(\mu\sigma)}=0.
\end{eqnarray}
If $\beta_{00}\neq0$, then taking into account first (\ref{P}) and then (\ref{M-sym}) we come to the
conclusion that $M_{\mu\nu\sigma}=M_{[\mu\nu\sigma]}=0$.
To avoid $M_{\mu\nu\sigma}=0$ we will adopt less strong condition that $\beta_{00}=0$.
In any case
this means that the background field $M_{[\mu\nu]\sigma}$ cannot influence
on the background geometry.

Now commutator $[l_1^+,l_0]$ takes the form
\begin{eqnarray}
[l_1^+,l_0]&\sim&
4a^{+\mu}a^{+\nu}a^\alpha R^\sigma{}_{\mu\nu\alpha}D_\sigma
+(2R^{\mu\sigma}+3\alpha_{00}F^{\sigma\mu})a_\mu^+ D_\sigma
-R_{\mu\alpha\beta\nu;\sigma}a^{\mu+}a^{\nu+}a^{\sigma+}a^\alpha a^\beta
\nonumber
\\
&&{}
+(4\alpha_{00}R^\sigma{}_{\mu\nu\alpha}V_\sigma-R_{\mu\nu;\alpha}
+\alpha_{00}\nabla_\mu F_{\nu\alpha})a^{+\mu}a^{\nu+}a^\alpha
\nonumber
\\
&&{}
+a^{+\mu}\alpha_{00}
\Bigl[
3\alpha_{00}F_{\sigma\mu}V^\sigma
+\nabla^\nu F_{\nu\mu}
+2V^\sigma R_{\sigma\mu}
\Bigr]
\,,
\label{l1+l0}
\end{eqnarray}
where $F_{\mu\nu}=\nabla_\mu V_\nu-\nabla_\nu V_\mu$.
We see that commutator (\ref{l1+l0}) does not proportional to the
operators (\ref{set-R}) if the curvature and the background vector field $V_\mu$
are arbitrary.
To find conditions on the curvature and $V_\mu$ which are necessary
for closing the algebra
we decompose the Riemann tensor into irreducible parts
\begin{eqnarray}
R_{\mu\nu\alpha\beta}&=&
C_{\mu\nu\alpha\beta}
+\frac{1}{d-2}\left(
\tilde{R}_{\mu\alpha}g_{\nu\beta}+\tilde{R}_{\nu\beta}g_{\mu\alpha}
-\tilde{R}_{\mu\beta}g_{\nu\alpha}-\tilde{R}_{\nu\alpha}g_{\mu\beta}
\right)
\nonumber
\\
&&{}
+\frac{R}{d(d-1)}(g_{\mu\alpha}g_{\nu\beta}-g_{\mu\beta}g_{\nu\alpha}),
\end{eqnarray}
where $C_{\mu\nu\alpha\beta}$ is the Weyl tensor,
$\tilde{R}_{\mu\nu}=R_{\mu\nu}-\frac{1}{d}g_{\mu\nu}R$ is the traceless part
of the Ricci tensor, and substitute this decomposition into (\ref{l1+l0})
\begin{eqnarray}
[l_1^+,l_0]&\sim&
4a^{+\mu}a^{+\nu}a^\alpha C^\sigma{}_{\mu\nu\alpha}D_\sigma
+3\alpha_{00}F^{\sigma\mu}a_\mu^+ D_\sigma
-C_{\mu\alpha\beta\nu;\sigma}a^{\mu+}a^{\nu+}a^{\sigma+}a^\alpha a^\beta
\nonumber
\\
&&{}
+\Bigl[\alpha_{00}4C^\sigma{}_{\mu\nu\alpha}V_\sigma
+\tilde{R}_{\mu\alpha;\nu}-\tilde{R}_{\mu\nu;\alpha}
+\alpha_{00}\nabla_\mu F_{\nu\alpha}
\Bigr]a^{+\mu}a^{\nu+}a^\alpha
\nonumber
\\
&&{}
+a^{+\mu}
\Bigl[
3\alpha_{00}^2F_{\sigma\mu}V^\sigma
+\alpha_{00}\nabla^\sigma F_{\sigma\mu}
-\frac{(d-2)(d-4)}{4d(d-1)}\nabla_\mu R
\Bigr]
.
\label{l1+l0-}
\end{eqnarray}
From first term of r.h.s. of (\ref{l1+l0-}) we find that it is necessary to suppose
\begin{eqnarray}
C_{\mu(\alpha\beta)\nu}=0\,,
&\qquad&
\alpha_{00}F_{\mu\nu}=\alpha_{00}(\nabla_\mu V_\nu-\nabla_\nu V_\mu)=0\,.
\label{cond1}
\end{eqnarray}
The left condition in (\ref{cond1}) together with the index symmetry
of the Weyl tensor tells us that the Weyl tensor is completely
antisymmetric $C_{\mu\alpha\beta\nu}=C_{[\mu\alpha\beta\nu]}$, and
due to the Bianchi identity $C_{\mu[\alpha\beta\nu]}=0$ it equals to
zero. To satisfy the right condition in (\ref{cond1}) we can put
$\alpha_{00}=0$ or $F_{\mu\nu}=0$, after that field $V_\mu$
disappears in the r.h.s. (\ref{l1+l0-}). In particular, if the
coefficient $\alpha_{00}$ is somehow fixed, e.g. the vector field
enters the constraints through the $U(1)$ covariant derivative, ones
get immediately that $F_{\mu\nu}=0$. This means in general that
under the above assumptions the vector field $V_\mu$ does not
influence on closing the algebra and as a consequence on the
background geometry.

Next, from the second line of (\ref{l1+l0-}) we find
\begin{eqnarray}
\tilde{R}_{\alpha(\mu;\nu)}=\tilde{R}_{\mu\nu;\alpha}
\label{cond2}
\end{eqnarray}
where we have used (\ref{cond1}).
Contracting indices $\mu$ and $\nu$ in (\ref{cond2}) and using the Bianchi
identity one gets
\begin{eqnarray}
\nabla_\mu R=0
&\Rightarrow&
R=const.
\label{cond2+}
\end{eqnarray}
As a result, the background geometry must be a constant curvature
space-time.

Thus we have shown that in the higher spin field theory the
background vector and third rank tensor cannot have influence on the
geometry of the space which must be only a constant curvature one.

Let us consider the case when the background fields are introduced
into the operators being multiplied on some operators of the
algebra, like, for example, they are introduced in the operator
$l_1$ (\ref{l1}), except the terms with coefficients $\alpha_{00}$,
$\beta_{00}$, $\gamma_{00}$. In this case we expect that the
background fields will make no effects, at least on the physical
field, the same as in quantization of gauge theories a redefinition
of constraints by terms proportional to constraints have no effect
on the physical states. In case of higher spin theory we illustrate
this on a simplified example.

Let the operators have the following form
\begin{align}
\label{EM-l0}
&l_0=\partial^2-m^2+\alpha_1 g_0+\alpha_2 g_0^2
\\
& l_1=a^\alpha \partial_\alpha && l_1^+=-a^{+\mu}\partial_\mu && g_m=m^2
\\
& l_2=\tfrac{1}{2}a_\mu a^\mu && l_2^+=\tfrac{1}{2}a_\mu^+ a^{+\mu}
&&g_0=a_\mu^+ a^\mu+\tfrac{d}{2}
\label{EM-g0}
\end{align}
where $\alpha_1$  and $\alpha_2$ are some combinations of the
background fields with the dimension of mass squared. Since
$\alpha_1$ and $\alpha_2$ are multiplied on operator of the algebra,
then the algebra is closed at any $\alpha_1$ and $\alpha_2$. To
simplify the subsequent calculations we adopt $\alpha_1$ and
$\alpha_2$ are constants. At first glance it seems that the mass
shell equation which we will reproduce using the BRST method must
will be $l_0|\Phi\rangle=0$, but as we shall show it will turn out
to be $(\partial^2-m^2)|\Phi\rangle=0$, thus removing all the
dependence on $\alpha$'s (this is what we mean in
footnote~\ref{fnote} at page~\pageref{fnote}). In case of constant
$\alpha$'s the algebra of operators (\ref{EM-l0})--(\ref{EM-g0}) has
the following non-vanishing commutators
\begin{align}
&[l_1^+,l_1]=l_0+g_m-\alpha_1g_0-\alpha_2g_0^2,
\\
&[l_0,l_1]=-2\alpha_2g_0l_1-(\alpha_1+\alpha_2)l_1,
&&[l_2^+,l_1]=l_1^+,
\\
&[l_0,l_1^+]=2\alpha_2l_1^+g_0+(\alpha_1+\alpha_2)l_1^+,
&&[l_1^+,l_2]=l_1,
\\
&[l_0,l_2]=-4\alpha_2g_0l_2-(2\alpha_1+4\alpha_2)l_2,
&&[l_2,l_2^+]=g_0,
\\
&[l_0,l_2^+]=4\alpha_2l_2^+g_0+(2\alpha_1+4\alpha_2)l_2^+,
&&[g_0,l_k^{\pm}]=\pm kl_k
.
\end{align}

According to the BRST method of Lagrangian construction (see e.g.
\cite{bkl}) since among the operators (\ref{EM-l0})--(\ref{EM-g0})
there are operators $g_0$, $g_m$ which are not constrains neither in
the space of bra vectors nor in the space of ket vectors then we
must construct extended expressions for the operators
$o_i\to{}O_i=o_i+o_i'$ where $o_i'$ are additional parts to the
initial operators $o_i=\{l_0,l_1,l_1^+l_2,l_2^+,g_0,g_m\}$
(\ref{EM-l0})--(\ref{EM-g0}). These additional parts are constructed
from new (additional) creation and annihilation operators and
commute with the initial operators $[o_i,o_j']=0$. The extended
expressions for the operators must satisfy two conditions: 1)~they
must form an algebra $[O_i,O_j]\sim{}O_k$; 2)~the operators which
are not constraints $g_0$, $g_m$ must be zero or contain linearly
arbitrary parameters which value will be defined later from the
condition of reproducing desired equations of motion.

Using the method elaborated in \cite{bkl} we find algebras of the additional
parts
\begin{align}
\label{EM-l1l1+'}
&
[l_1',l_1^{\prime+}]=-l_0'-g_m'+\alpha_1g_0'-\alpha_2g_0^{\prime2},
\\
&
\label{EM-l1l0'}
[l_1',l_0']=-2\alpha_2g_0'l_1'+(\alpha_1-\alpha_2)l_1',
&&
[l_2^{\prime+},l_1']=l_1^{\prime+},
\\
\label{EM-l0l1+'}
&
[l_0',l_1^{+\prime}]=-2\alpha_2l_1^{+\prime}g_0'+(\alpha_1-\alpha_2)l_1^{+\prime},
&&
[l_1^{\prime}+,l_2']=l_1',
\\
\label{EM-l0l2'}
&
[l_2',l_0']=-4\alpha_2g_0'l_2'+(2\alpha_1-4\alpha_2)l_2',
&&
[l_2',l_2^{\prime+}]=g_0',
\\
\label{EM-l0l2+'}
&
[l_0',l_2^{+\prime}]=-4\alpha_2l_2^{+\prime}g_0'+(2\alpha_1-4\alpha_2)l_2^{+\prime},
&&
[g_0',l_k^{\prime\pm}]=\pm kl_k^{\prime\pm}
\end{align}
and of the extended operators
\begin{align}
&
\label{EM-L1+L1}
[L_1^+,L_1]=L_0+G_m-\alpha_1G_0-\alpha_2G_0^2+2\alpha_2g_0'G_0
\\{}
\label{EM-L1L0}
&
[L_1,L_0]=\alpha_2(G_0L_1+L_1G_0)+\alpha_1L_1
-2\alpha_2g_0'L_1-2\alpha_2l_1'G_0,
&&
[L_2^+,L_1]=L_1^+,
\\{}
\label{EM-L0L1+}
&
[L_0,L_1^+]=\alpha_2(L_1^+G_0+G_0L_1^+)+\alpha_1L_1^+
-2\alpha_2g_0'L_1^+-2\alpha_2l_1^{\prime+}G_0,
&&
[L_1^+,L_2]=L_1,
\\{}
\label{EM-L0L2}
&
[L_2,L_0]=2\alpha_2(G_0L_2+L_2G_0)+2\alpha_1L_2
-4\alpha_2g_0'L_2-4\alpha_2l_2'G_0,
&&
[L_2,L_2^+]=G_0,
\\{}
\label{EM-L0L2+}
&
[L_0,L_2^+]=2(L_2^+G_0+G_0L_1^+)+2\alpha_1L_2^+
-4\alpha_2g_0'L_2^+-4\alpha_2l_2^{+\prime}G_0,
&&
[G_0,L_k^{\pm}]=\pm kL_k^{\pm}.
\end{align}
In RHS of (\ref{EM-L1+L1})--(\ref{EM-L0L2+}) we choose symmetric odering of
the extended operators.
There is a method allowing to construct explicit form of operators in terms
of creation and annihilation operators on the base of their algebra, see e.g.
\cite{bkl} and references therein.
But for our purpose we need no any explicit realization of the additional
parts, except only one observation
\begin{eqnarray}
l_0'=\alpha_1g_0'-\alpha_2g_0^{\prime2}
.
\label{EM-l0'}
\end{eqnarray}

The BRST operator constructed on the base of the algebra of the
extended operators is
\begin{eqnarray}
\tilde{Q}
&=&
\eta_0L_{0}+\eta_1^+L_1+\eta_1L_1^++\eta_2^+L_2+\eta_2L_2^++\eta_{G}G_0
+\eta_MG_m
+\eta_1^+\eta_1({\cal{}P}_0+\mathcal{P}_M)
\nonumber
\\&&
{}
+(\eta_G\eta_1^++\eta_2^+\eta_1){\cal{}P}_1
+(\eta_1\eta_G+\eta_1^+\eta_2){\cal{}P}_1^+
+2\eta_G\eta_2^+{\cal{}P}_2
+2\eta_2\eta_G{\cal{}P}_2^+
\nonumber
\\&&
{}
-\eta_2^+\eta_2{\cal{}P}_G
-\eta_1^+\eta_1\Bigl[\alpha_1+\alpha_2(G_0-2g_0')\Bigr]\mathcal{P}_G
\nonumber
\\&&
{}
+\alpha_2\eta_0\Bigl[
\eta_1^+(L_1-2l_1')-\eta_1(L_1^+-2l_1^{+\prime})
+2\eta_2^+(L_2-2l_2')-2\eta_2(L_2^+-2l_2^{+\prime})
\Bigr]
\mathcal{P}_G
\nonumber
\\&&
{}
+\eta_0\Bigl[\alpha_1+\alpha_2(G_0-2g_0')\Bigr]\Bigl(
\eta_1^+\mathcal{P}_1-\eta_1\mathcal{P}_1^++2\eta_2^+\mathcal{P}_2-2\eta_2\mathcal{P}_2^+
\Bigr)
,
\label{EM-BRST}
\end{eqnarray}

Next step of Lagrangian construction is determination of the
arbitrary parameters which must be contained linearly in the
additional parts $g_0'$ and $g_m'$. For this we decompose the BRST
operator extracting its dependence on ghosts $\eta_G$,
$\mathcal{P}_G$, $\eta_M$, $\mathcal{P}_M$, corresponding to these
operators,
\begin{eqnarray}
\tilde{Q}&=&Q+\eta_G\tilde{G}_0+\eta_MG_M+\eta_1^+\eta_1\mathcal{P}_M+\mathcal{B}\mathcal{P}_G
\end{eqnarray}
where
\begin{eqnarray}
Q
&=&
\eta_0L_{0}+\eta_1^+L_1+\eta_1L_1^++\eta_2^+L_2+\eta_2L_2^+
+\eta_1^+\eta_1{\cal{}P}_0
+\eta_2^+\eta_1{\cal{}P}_1
+\eta_1^+\eta_2{\cal{}P}_1^+
\nonumber
\\&&
{}
+\eta_0\Bigl[\alpha_1+\alpha_2(G_0-2g_0')\Bigr]\Bigl(
\eta_1^+\mathcal{P}_1-\eta_1\mathcal{P}_1^++2\eta_2^+\mathcal{P}_2-2\eta_2\mathcal{P}_2^+
\Bigr)
,
\label{EM-Q}
\\
\tilde{G}_0&=&G_0+
\eta_1^+\mathcal{P}_1-\eta_1\mathcal{P}_1^++2\eta_2^+\mathcal{P}_2-2\eta_2\mathcal{P}_2^+
,
\end{eqnarray}
(explicit expression for $\mathcal{B}$ is not essential)
and suppose that the state vector $|\Psi\rangle$ in the extended space including ghosts
does not depend on ghosts $\eta_G$ and $\eta_M$,
$\mathcal{P}_G|\Psi\rangle=\mathcal{P}_M|\Psi\rangle=0$.
As a result the equation defining physical states $\tilde{Q}|\Psi\rangle=0$
is decomposed into three equations
\begin{align}
&Q|\Psi\rangle=0,
&&\tilde{G}_0|\Psi\rangle=0,
&&G_M|\Psi\rangle=0.
\label{EM-3}
\end{align}
Two right equations in (\ref{EM-3}) are used for determination of the
arbitrary constants in $g_0'$ and $g_m'$ and the left equation in
(\ref{EM-3}) is the equation on physical states.
Note that using (\ref{EM-l0'}) and $L_0=l_0+l_0'$ where $l_0$ is given by
(\ref{EM-l0}) operator $Q$ (\ref{EM-Q}) can be rewritten as
\begin{eqnarray}
Q
&=&
\eta_0\Bigl[\partial^2-m^2+\alpha_1\tilde{G}_0+\alpha_2(G_0-2g_0')\tilde{G}_0\Bigr]
\nonumber
\\&&
{}
+\eta_1^+L_1+\eta_1L_1^++\eta_2^+L_2+\eta_2L_2^+
+\eta_1^+\eta_1{\cal{}P}_0
+\eta_2^+\eta_1{\cal{}P}_1
+\eta_1^+\eta_2{\cal{}P}_1^+
\label{EM-QQ}
\end{eqnarray}
and due to the middle equation in (\ref{EM-3}) all the effects of
the ``constant background fields'' $\alpha_1$ and $\alpha_2$
disappear and we get the model of free higher spin field in
Minkowski space.


To summarize, we have developed the BRST approach to Lagrangian
construction for bosonic totally symmetric higher spin field in
external gravitational, vector and third rank tensor fields.
Assuming that interaction with external fields has massless and flat
space limits we prove that the consistent formulation is possible
only in constant curvature space with no nontrivial coupling to the
third rank tensor and the strength of the vector fields. One can
expect that analogous situation will take place for higher spin
fermionic fields and for any deformation of constant curvature space
by more general background tensor fields. However, the above result
does not concern the field models with spins $\frac{3}{2}$ and $2$
where the BRST construction has the specific possibilities
\cite{brst1} and allows the consistent Lagrangian formulation in
Einstein spaces. Thus, the further development of Lagrangian
construction for free higher spin fields interacting with external
fields is related to search for interaction Lagrangians which
contain the inverse mass terms. Some approaches to such Lagrangians
are given in refs. \cite{consist1,consist2}. Also, it would be
interesting to study the consistency conditions for for recently
formulated conformal higher spin fields \cite{Metsaev} if to couple
them to external fields.


{\bf {Acknowledgements}}. The authors are grateful to Yu.M. Zinoviev
and R.R. Metsaev for discussions of some aspects. The work is
partially supported by the RFBR grant, project No.\ 09-02-00078,
grant for LRSS, project No.\ 3558.2010.2. The work of I.L.B. and
P.M.L. is also partially supported by the RFBR-DFG grant, project
No.\ 09-02-91349 and the DFG grant, project No. 436 RUS 113/669/0-4.


\end{document}